\documentclass[pra,twocolumn,preprintnumbers,showpacs]{revtex4}
\usepackage{amsmath,amssymb,bm,graphicx}

\newcommand\eps{\epsilon}
\renewcommand\d{\partial}
\newcommand\grad{\bm{\nabla}}
\newcommand\+{\dagger}

\newcommand\p{{\bm{p}}}
\newcommand\q{{\bm{q}}}
\renewcommand\k{{\bm{k}}}
\newcommand\ep{{\varepsilon_\p}}
\newcommand\eq{{\varepsilon_\q}}
\newcommand\ek{{\varepsilon_\k}}
\newcommand\eF{\varepsilon_\mathrm{F}}

\begin{document}
\preprint{INT-PUB 08-43}

\title{Ground-state energy of the unitary Fermi gas from the $\epsilon$ expansion}
\author{Yusuke~Nishida}
%\email{nishida@phys.washington.edu}
%\homepage{http://tkynt2.phys.s.u-tokyo.ac.jp/~nishida/}
\affiliation{Institute for Nuclear Theory, University of Washington,
             Seattle, Washington 98195-1550, USA}

\begin{abstract}
 We update the ground-state energy ratio of unitary Fermi gas to
 noninteracting Fermi gas ($\xi$) from the $\epsilon$ expansion by
 including the next-to-next-to-leading-order (NNLO) term near two
 spatial dimensions.  Interpolations of the NNLO $\epsilon$ expansions
 around four and two spatial dimensions with the use of Pad\'e
 approximants give $\xi\approx0.360\pm0.020$ in three dimensions with
 the uncertainty due to different interpolation functions.  This value
 is consistent with the previous interpolations of the NLO $\epsilon$
 expansions $\xi\approx0.377\pm0.014$ in spite of the large NNLO
 corrections.
\end{abstract}

\date{August 2008}
%03.75.Ss Degenerate Fermi gases
%05.30.Fk Fermion systems and electron gas
%67.85.Lm Degenerate Fermi gases

\maketitle

\section{Introduction}
Two-component fermions interacting via a zero-range and infinite
scattering length interaction have attracted intense attention across
many subfields of physics~\cite{review_theory}.  Experimentally, such a
system can be realized in trapped atoms using the Feshbach resonance and
has been extensively studied~\cite{Ketterle:2008}.  The most important
property of the system is the scale invariance of the interaction, and
thus, it can be thought of a rare realization of nonrelativistic
conformal field
theories~\cite{Mehen:1999nd,Son:2005rv,Nishida:2007pj,Mehen:2007dn}.

As a consequence of the scale invariance of the interaction, all
physical quantities at finite density and zero temperature are
determined by simple dimensional analysis up to dimensionless constants
of proportionality.  Such dimensionless parameters are universal
depending only on the dimensionality of space.  A representative example
of the universal parameters is the ground-state energy of the Fermi gas
at infinite scattering length (unitary Fermi gas) normalized by that of
a noninteracting Fermi gas with the same density:
\begin{equation}\label{eq:xi}
 \xi_d \equiv \frac{E_\mathrm{unitary}}{E_\mathrm{free}}.
\end{equation}
Here we put a subscript $d$ to emphasize that $\xi_d$ is a function of
the dimensionality of space.  Because $\xi_d$ is a fundamental quantity
characterizing the unitary Fermi gas, there have been substantial
efforts to determine its value in $d=3$ both from
experiments~\cite{OHara:2002,Bartenstein:2004,Bourdel:2004,Kinast:2005,Partridge:2006,Stewart:2006,Tarruell:2007}
and Monte Carlo
simulations~\cite{Carlson:2003,Chang:2004,Astrakharchik:2004,Carlson:2005,Bulgac:2006,Lee:2006,Abe:2007,Bulgac:2008,Lee:2008,Zhang}.

For analytical treatments, the scale-invariant interaction implies great
difficulties because there seems to be no parameter to control a theory.
However, it was shown that the problem of unitary Fermi gas can be
solved systematically with appropriately formulated perturbation
theories if the dimensionality of space $d$ is close to 4 or close to
2~\cite{Nishida:2006br,Nishida:2006eu,Nishida:2006rp}.  This is inspired
by the special nature of four and two spatial dimensions for the
zero-range and infinite scattering length
interaction~\cite{Nussinov:2006}: the unitary Fermi gas becomes a
noninteracting Bose gas in $d=4$ ($\xi_{d\to4}\to0$), while it becomes a
noninteracting Fermi gas in $d=2$ ($\xi_{d\to2}\to1$).  Corrections to
$\xi_d$ near four and two spatial dimensions have been computed up to
next-to-next-to-leading order (NNLO) in terms of $\eps=4-d$ and
$\bar\eps=d-2$~\cite{Nishida:2006br,Nishida:2006eu,Arnold:2006fr,Nishida:2006wk}:
\begin{align}
 \label{eq:xi_4d}
 \xi_{4-\eps} &= \frac{\eps^{(6-\eps)/(4-\eps)}}2 \\
 &\times \left[1 - 0.04916\,\eps
 - 0.95961\,\eps^2 + O(\eps^3)\right] \notag\\
 \intertext{and}
 \xi_{2+\bar\eps}
 &= 1 - \bar\eps + 0.80685\,\bar\eps^2 + O(\bar\eps^3). 
 \label{eq:xi_2d}
\end{align}
Because NNLO corrections turn out to be large, naive extrapolations of
the $\eps$ and $\bar\eps$ expansions to the physical case in $d=3$ do
not work at all. The more appropriate way to obtain the value of $\xi_d$
in $d=3$ is to interpolate the two expansions.  This procedure has been
carried out by using the next-to-leading-order (NLO) expansions around
$d=4$ and $d=2$~\cite{Nishida:2006eu} and by using the NNLO expansion
around $d=4$ and the NLO expansion around $d=2$~\cite{Arnold:2006fr},
and reasonable agreement with results from Monte Carlo simulations was
found.

The main purpose of this paper is to update $\xi_d$ in $d=3$ by
including the NNLO term near two spatial dimensions.  First, we review
the interpolations of the NLO $\eps$ expansions to see the stability of
the results to the choice of interpolation schemes (Sec.~\ref{sec:NLO}).
We then show results from the interpolations of the NNLO $\eps$
expansions in Sec.~\ref{sec:NNLO}.  Finally, a summary and concluding
remarks are given in Sec.~\ref{sec:summary}.  The NNLO correction to
$\xi_d$ near $d=2$ shown in Eq.~(\ref{eq:xi_2d}) is computed in the
Appendix.

\section{Interpolations of NLO expansions \label{sec:NLO}}
In order to see the stability of the results to the choice of
interpolation schemes, we review the interpolations of the NLO $\eps$
expansions by using Pad\'e approximants with and without applying the
Borel transformation.  

\begin{figure*}[tp]\hfill
 \includegraphics[width=0.45\textwidth,clip]{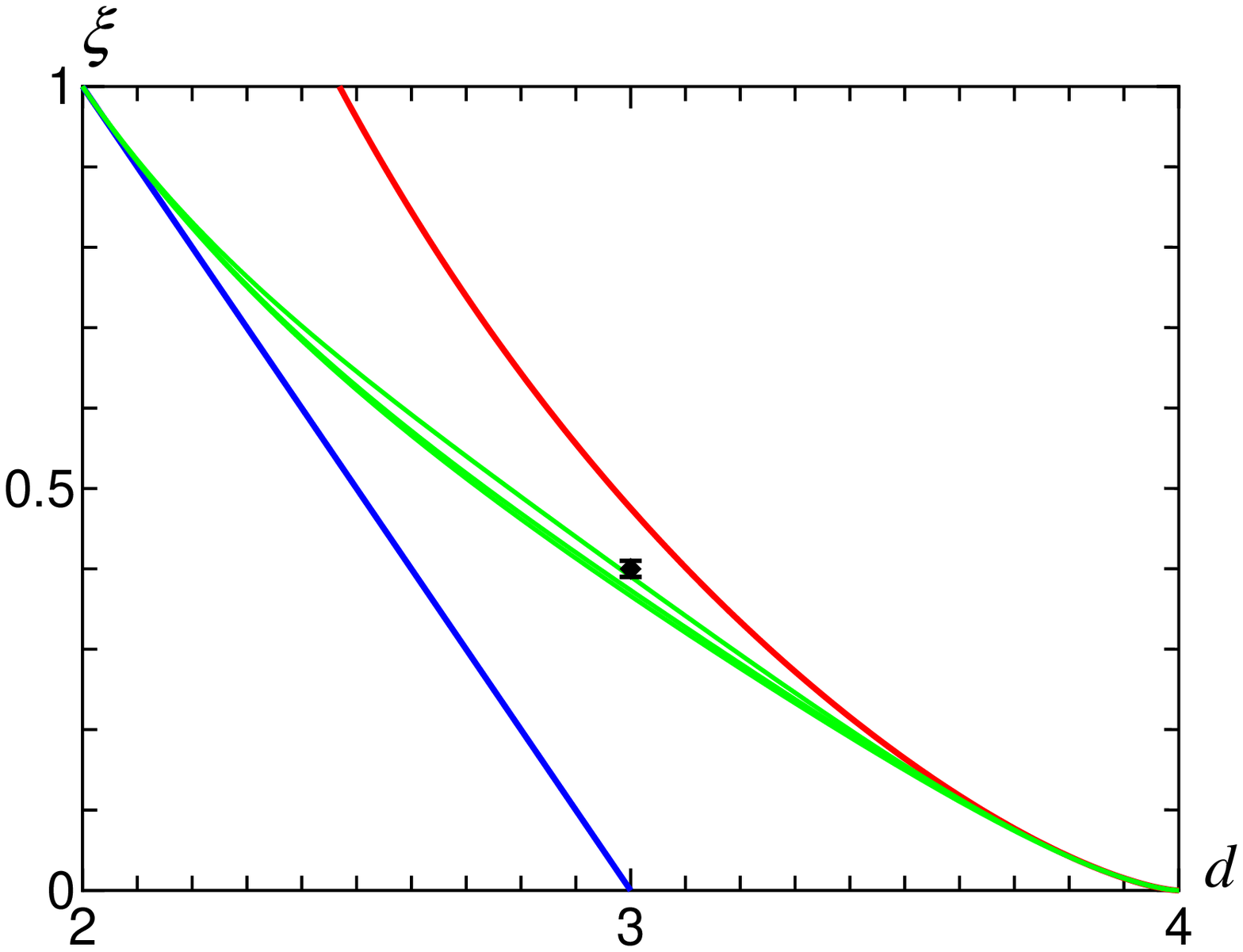}\hfill
 \includegraphics[width=0.45\textwidth,clip]{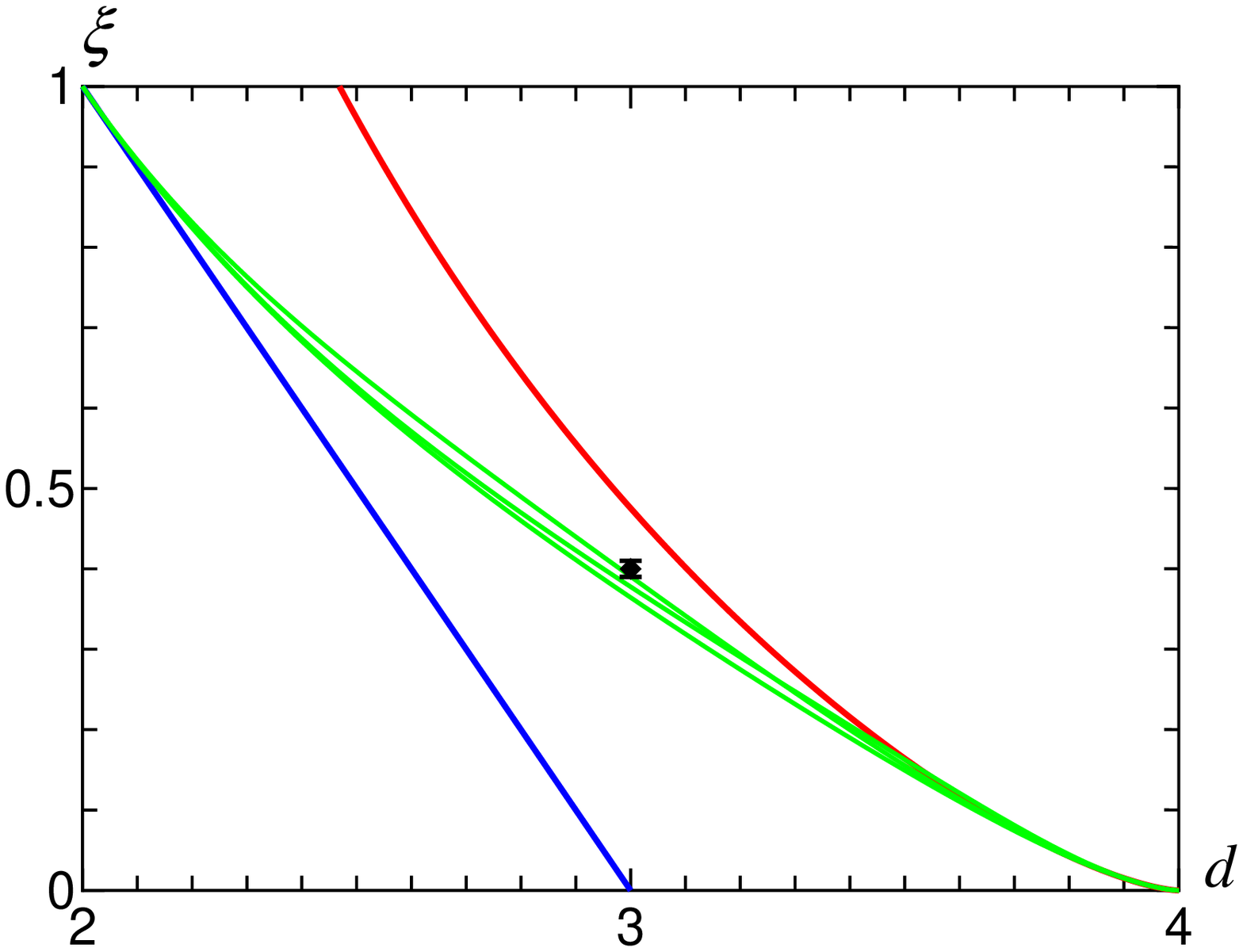}\hfill\hfill
 \caption{(Color online) The universal parameter $\xi_d$ as a function
 of spatial dimensions $d$.  The upper curve is the extrapolation from
 the NLO expansion around $d=4$ in Eq.~(\ref{eq:xi_4d}), while the lower
 line is the extrapolation from the NLO expansion around $d=2$ in
 Eq.~(\ref{eq:xi_2d}).  The middle three curves show the Pad\'e (left
 panel) and Borel-Pad\'e~\cite{Nishida:2006eu} (right panel)
 interpolations of the two NLO expansions.  The symbol at $d=3$
 indicates the result $\xi_3\approx0.40(1)$ from the latest Monte Carlo
 simulations~\cite{Bulgac:2008,Zhang}.  \label{fig:xi_nlo}}
\end{figure*}

\subsection{Pad\'e interpolation}
The simplest way to interpolate the two expansions around $d=4$ and
$d=2$ is to use the Pad\'e approximants.  We write $\xi_d$ in
Eq.~(\ref{eq:xi_4d}) in the following form: 
\begin{equation}\label{eq:F}
  \xi_{4-\eps} = \frac{\eps^{(6-\eps)/(4-\eps)}}2 F(\eps),
\end{equation}
where $F(\eps)$ is an unknown function having the expansion
$F(\eps)=1-0.04916\,\eps-0.95961\,\eps^2+O(\eps^3)$~\footnote{It has
been shown that there is a nonanalytic term $-\frac38\eps^3\ln\eps$ to the
next-to-next-to-next-to-leading order in $\eps$~\cite{Arnold:2006fr}.
Because we are working up to NNLO in the current paper, we neglect such
a nonanalytic contribution.}.  We approximate $F(\eps)$ by a ratio of
two polynomials (Pad\'e approximant),
\begin{equation}
 F_{[M/N]}(\eps) = \frac{p_0+p_1\eps+\cdots+p_M\eps^M}
  {1+q_1\eps+\cdots+q_N\eps^N},
\end{equation}
and determine the unknown coefficients so that $\xi_d$ has the correct
expansions around $d=4$ and $d=2$.  If one truncates the $\eps$ and
$\bar\eps$ expansions at NLO, we have four known terms and thus Pad\'e
approximants $F_{[M/N]}$ satisfying $M+N=3$ are possible.  We exclude
the possibility of $F_{[2/1]}(\eps)$ because it has a pole in a range
$0<\eps<2$, while we expect a smooth behavior of $\xi_d$ as a function
of $2<d<4$.

The left panel in Fig.~\ref{fig:xi_nlo} shows the universal parameter
$\xi_d$ as a function of $d$.  The middle three curves show the Pad\'e
interpolations of the two NLO expansions with the use of $F_{[3/0]}$,
$F_{[1/2]}$, and $F_{[0/3]}$.  In $d=3$, these interpolations,
respectively, give
\begin{equation}\label{eq:nlo_pade}
 \xi_3 \approx 0.391,\ \ 0.366,\ \ 0.373.
  %0.390516 [3/0], 0.366359 [1/2], 0.373412 [0/3]
  %0.376762\pm0.0137537
\end{equation}
These three values have an average $0.377$ and span a small interval
$\xi_3\approx0.377\pm0.014$.  We note that the same interpolation scheme
was employed to compute the lowest two energy levels of three fermions
in a harmonic potential, and excellent agreement with the exact results
was found in arbitrary spatial dimensions
$2<d<4$~\cite{Nishida:2007pj}.

\subsection{Borel-Pad\'e interpolation}
The other way to interpolate the two expansions is to apply the Borel
transformation and then use the Pad\'e
approximants~\cite{Nishida:2006eu}.  We first rewrite the unknown
function $F(\eps)$ in Eq.~(\ref{eq:F}) in the form of the Borel
transformation:
\begin{equation}
 F(\eps) = \frac1{\eps}\int_0^\infty\!dt\,e^{-t/\eps}G(t).
\end{equation}
If $F(\eps)$ has an expansion $F(\eps)=\sum_{n=0}^{\infty}c_n\eps^n$,
the Borel transform $G(t)$ has an expansion
$G(t)=\sum_{n=0}^{\infty}\frac{c_n}{n!}t^n$, and thus, the Borel
transformation makes the expansion faster convergent.  Then we
approximate $G(t)$ by the Pad\'e approximant,
\begin{equation}
 G_{[M/N]}(t) = \frac{p_0+p_1t+\cdots+p_Mt^M}{1+q_1t+\cdots+q_Nt^N},
\end{equation}
and determine the unknown coefficients so that $\xi_d$ has the correct
expansions around $d=4$ and $d=2$.

The right panel in Fig.~\ref{fig:xi_nlo} shows the universal parameter
$\xi_d$ as a function of $d$.  The middle three curves show the
Borel-Pad\'e interpolations of the two NLO expansions with the use of
$G_{[3/0]}$, $G_{[1/2]}$, and $G_{[0/3]}$.  The possibility of
$G_{[2/1]}$ is excluded because we could not find a solution satisfying
the constraints of Eqs.~(\ref{eq:xi_4d}) and (\ref{eq:xi_2d}).  In
$d=3$, these interpolations, respectively, give~\cite{Nishida:2006eu}
\begin{equation}\label{eq:nlo_borel-pade}
 \xi_3 \approx 0.391,\ \ 0.364,\ \ 0.378.
  %0.390516 [3/0], 0.364396 [1/2], 0.377749 [0/3]
  %0.377554\pm0.0131577
\end{equation}
These three values have an average $0.378$ and span a small interval
$\xi_3\approx0.378\pm0.013$.  We note that the result of $G_{[3/0]}$ is
equivalent to that of $F_{[3/0]}$ in Eq.~(\ref{eq:nlo_pade}).

Comparing the results in Eqs.~(\ref{eq:nlo_pade}) and
(\ref{eq:nlo_borel-pade}), one can see that the interpolated values do
not depend much on the choice of the Pad\'e approximants and also the
Borel transformation does not improve the interpolated values.  The
latter can be understood because we have only a few terms in the
expansion over $\eps$ (two terms up to NLO), and thus the advantage to
apply the Borel transformation is little.  This situation does not
change even if we include the NNLO term near $d=4$.  The use of the
Borel transformation may become important once we have more higher-order
corrections and the $\eps$ expansion is not a convergent series.

\section{Pad\'e interpolation of NNLO expansions \label{sec:NNLO}}
We now include the NNLO terms near $d=4$ and $d=2$ to interpolate the
two expansions.  Here we only use the Pad\'e interpolation by the
above-mentioned reason~\footnote{Of course one may apply the Borel
transformation to the NNLO $\eps$ expansion near $d=4$ and then
approximate the Borel transform $G(t)$ by some interpolation functions.
If we use the Pad\'e approximants as the interpolation functions, we
find a nontrivial solution satisfying the constraints of
Eqs.~(\ref{eq:xi_4d}) and (\ref{eq:xi_2d}) only in $G_{[4/1]}$, which
gives $\xi_3\approx0.373$ in $d=3$.  %0.372922
Note that the result of $G_{[5/0]}$ is trivially equivalent to that of
$F_{[5/0]}$ in Eq.~(\ref{eq:nnlo_pade}).}.  Because we have six known
terms, Pad\'e approximants $F_{[M/N]}$ satisfying $M+N=5$ are possible.
However, we exclude the possibility of $F_{[2/3]}$ and $F_{[1/4]}$
because they have poles in a range $0<\eps<2$, while we expect a smooth
behavior of $\xi_d$ as a function of $2<d<4$.

\begin{figure}[tp]
 \includegraphics[width=0.45\textwidth,clip]{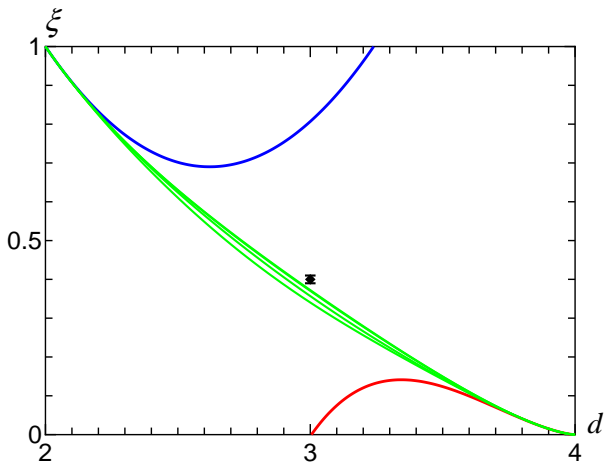}
 \caption{(Color online) The universal parameter $\xi_d$ as a function
 of spatial dimensions $d$.  The lower right curve is the extrapolation
 from the NNLO expansion around $d=4$ in Eq.~(\ref{eq:xi_4d}), while the
 upper left curve is the extrapolation from the NNLO expansion around
 $d=2$ in Eq.~(\ref{eq:xi_2d}).  The middle four curves show the Pad\'e
 interpolations of the two NNLO expansions.  The symbol at $d=3$
 indicates the result $\xi_3\approx0.40(1)$ from the latest Monte Carlo
 simulations~\cite{Bulgac:2008,Zhang}.  \label{fig:xi_nnlo}}
\end{figure}

Figure~\ref{fig:xi_nnlo} shows the universal parameter $\xi_d$ as a
function of $d$.  The middle four curves show the Pad\'e interpolations
of the two NNLO expansions with the use of $F_{[5/0]}$, $F_{[4/1]}$,
$F_{[3/2]}$, and $F_{[0/5]}$.  In $d=3$, these interpolations,
respectively, give
\begin{equation}\label{eq:nnlo_pade}
 \xi_3 \approx 0.340,\ \ 0.372,\ \ 0.370,\ \ 0.357.
  %0.340470 [5/0], 0.372450 [4/1], 0.370415 [3/2], 0.356982 [0/5]
  %0.360079\pm0.0196092
\end{equation}
These four values have an average $0.360$ and span an interval
$\xi_3\approx0.360\pm0.020$~\footnote{If we neglected the interpolation
by the simple polynomial $F_{[5/0]}$ as was done in
Ref.~\cite{Arnold:2006fr}, we would obtain $0.367\pm0.010$.
%0.366616\pm0.00963340
This value is consistent with the Borel-Pad\'e interpolations without
the NNLO correction near $d=2$; $0.367\pm0.009$~\cite{Arnold:2006fr}.}.
%0.367393\pm0.00859399
It is understandable that the interpolations of the NNLO expansions have
the larger uncertainty because of the large NNLO corrections both near
$d=4$ and $d=2$ [see Eqs.~(\ref{eq:xi_4d}) and (\ref{eq:xi_2d}) and also
Fig.~\ref{fig:xi_nnlo}].  What is remarkable is that in spite of such
large NNLO corrections, the interpolated values are consistent with the
previous interpolations of the NLO expansions
$\xi_3\approx0.377\pm0.014$.  Therefore we conclude that the
interpolated results are stable to inclusion of higher-order corrections
and thus the $\eps$ expansion has a certain predictive power even though
the knowledge on higher-order terms in the expansions over $\eps=4-d$
and $\bar\eps=d-2$ is currently lacking.

\section{Summary and concluding remarks \label{sec:summary}}
In this paper, we have updated the ground-state energy ratio of unitary
Fermi gas to noninteracting Fermi gas ($\xi$) from the $\eps$ expansion
by including the NNLO term near two spatial dimensions.  We found that
the Pad\'e interpolations of the NNLO expansions around $d=4$ and $d=2$
give $\xi\approx0.360\pm0.020$ in $d=3$ with the relatively small
uncertainty from different interpolation functions.  Although the NNLO
corrections are large both near $d=4$ and $d=2$, the interpolated value
is consistent with the interpolations of the NLO expansions
$\xi\approx0.377\pm0.014$.  This indicates that the interpolated results
are stable to inclusion of higher-order corrections, and thus the $\eps$
expansion has a certain predictive power.  Indeed, our interpolated
values reasonably agree with the results from the latest Monte Carlo
simulations, $\xi\approx0.40(5)$~\cite{Bulgac:2008} and
$\xi\lesssim0.40(1)$~\cite{Zhang}.

Our analysis also implies that in order to obtain appropriate results
from the $\eps$ expansion, it is necessary to incorporate the expansions
both around $d=4$ and $d=2$.  Other than $\xi$ studied in this paper,
interpolations of NLO expansions around $d=4$ and $d=2$ have been
employed to estimate the critical temperature
$T_\mathrm{c}$~\cite{Nishida:2006rp}, thermodynamic functions at
$T_\mathrm{c}$~\cite{Nishida:2006rp}, and the ground-state energy of a
few fermions in a harmonic potential~\cite{Nishida:2007pj}.
Quasiparticle spectrum~\cite{Nishida:2006br,Nishida:2006eu}, atom-dimer
and dimer-dimer scatterings in vacuum~\cite{Rupak:2006jj}, the phase
structure of polarized Fermi gas with equal
masses~\cite{Nishida:2006eu,Rupak:2006et} and unequal
masses~\cite{Nishida:2006wk}, BCS-BEC crossover~\cite{Chen:2006wx},
momentum distribution and condensate fraction~\cite{Nishida:2006wk},
low-energy dynamics~\cite{Kryjevski:2007au}, and energy-density
functional~\cite{Rupak:2008xq} have been studied only in the expansions
over $\eps=4-d$.  It is possible to obtain better understanding of these
subjects by further incorporating the expansions in terms of
$\bar\eps=d-2$.

\acknowledgments
The author thanks D.~Gazit for discussions.  This work was supported, in
part, by JSPS Postdoctoral Program for Research Abroad.

\appendix*
\section{NNLO correction to $\xi_d$ near $d=2$}
In this appendix, we briefly review the $\bar\eps$ expansion for the
unitary Fermi gas around two spatial dimensions and compute the NNLO
correction to $\xi_d$ in terms of $\bar\eps=d-2$ shown in
Eq.~(\ref{eq:xi_2d}).  The detailed account of the $\bar\eps$ expansion
is found in Ref.~\cite{Nishida:2006eu}.

\subsection{Lagrangian and power counting rule of $\bar\eps$}
The unitary Fermi gas near two spatial dimensions is described by the
sum of following Lagrangian densities (here and below $\hbar=1$):
\begin{align}
 \mathcal{L}_0 &= \sum_{\sigma=\uparrow,\downarrow}
 \psi_\sigma^\+\left(i\d_t+\frac{\grad^2}{2m}+\mu\right)\psi_\sigma, \\
 \mathcal{L}_1 & = -\varphi^*\varphi 
 + \bar g\varphi^*\psi_\downarrow\psi_\uparrow
 + \bar g\psi_\uparrow^\+\psi_\downarrow^\+\varphi, \\
 \mathcal{L}_2 & = \varphi^*\varphi.
\end{align}
Here we have neglected the condensate $\phi_0\sim\mu\,e^{-1/\bar\eps}$,
because its contribution is negligible compared to any power corrections
of $\bar\eps$.

The first part $\mathcal{L}_0$ generates the propagator of fermionic
field $\psi_\sigma$,
\begin{equation}
 G(p_0,\p) = \frac1{p_0-\ep+\mu+i\delta},
\end{equation}
where $\ep=\p^2/(2m)$ is the kinetic energy of nonrelativistic
particles.  The second part $\mathcal{L}_1$ describes the interaction
between fermions mediated by the auxiliary field $\varphi$.  The first
term in $\mathcal{L}_1$ gives the propagator of $\varphi$,
\begin{equation}
 D(p_0,\p) = -1, 
\end{equation}
and the last two terms give vertices coupling two fermions with
$\varphi$.  The coupling constant $\bar g$ is given by
\begin{equation}\label{eq:coupling-2d}
 \bar g = \left(\frac{2\pi\bar\eps}m\right)^{1/2}
  \left(\frac{m\mu}{2\pi}\right)^{-\bar\eps/4}.
\end{equation}
Here the factor $\left(m\mu/2\pi\right)^{-\bar\eps/4}$ was introduced so
that the product of auxiliary fields $\varphi^*\varphi$ has the same
dimension as the Lagrangian density.  We emphasize that the choice of
this factor is arbitrary, if it has the correct dimension, and does not
affect final results because the difference can be absorbed by the
redefinition of $\varphi$.  The particular choice of $\bar g$ in
Eq.~(\ref{eq:coupling-2d}) will simplify expressions for loop integrals in
intermediate steps. 

If we did not have the last part $\mathcal{L}_2$, we could integrate out
the auxiliary fields $\varphi$ and $\varphi^*$,which leads to
\begin{equation}
 \mathcal{L}_1\to \bar g^2\psi^\dagger_\uparrow\psi^\dagger_\downarrow
  \psi_\downarrow\psi_\uparrow,
\end{equation}
which represents the contact interaction of fermions with the small
coupling $\bar g^2\sim\bar\eps$.   Therefore, the unitary Fermi gas near
two spatial dimensions is simply described by a weakly interacting
system of fermions.  The vertex in $\mathcal{L}_2$ plays a role of a
counterterm so as to avoid double counting of a certain type of diagrams
which is already taken into $\mathcal{L}_1$.

\begin{figure}[tp]
 \includegraphics[width=0.48\textwidth,clip]{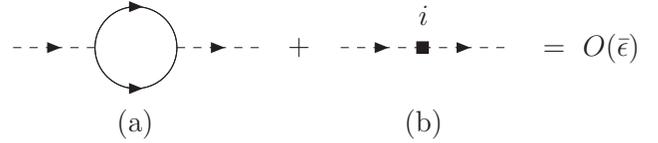}
 \caption{Power counting rule of $\bar\eps$.  The self-energy diagram of
 $\varphi$ field (a) is combined with the vertex from
 $\mathcal{L}_2$ (b) to achieve the simple $\bar\eps$ counting.  Solid
 (dotted) lines represent the fermion (auxiliary field) propagators $iG$
 ($iD$).  \label{fig:cancel-2d}}
\end{figure}

The power counting rule of $\bar\eps$ is summarized as follows.
\begin{enumerate}
 \item For any Green's function, we write down all Feynman diagrams
       using the propagator from $\mathcal{L}_0$ and the vertices from
       $\mathcal{L}_1$.
 \item If there is any subdiagram of the type in
       Fig.~\ref{fig:cancel-2d}(a), we add the same Feynman diagram where
       the subdiagram is replaced by the vertex from $\mathcal{L}_2$ in
       Fig.~\ref{fig:cancel-2d}(b).
 \item The power of $\bar\eps$ for the given Feynman diagram is simply
       $O\!\left(\bar\eps^{N_{\bar g}/2}\right)$, where $N_{\bar g}$ is
       the number of couplings $\bar g$.
\end{enumerate}
Here the dimensional regularization of loop integrals is assumed.

\subsection{Computation of the pressure}
The pressure of unitary Fermi gas has been computed up to the
next-to-leading order in $\bar\eps$~\cite{Nishida:2006eu}.  To the
leading order, the pressure is given by that of noninteracting fermions:
\begin{equation}
 \begin{split}
  P_\mathrm{free} &= 2\int\!\frac{d\p}{(2\pi)^d} 
  (\mu-\ep)\,\theta(\mu-\ep) \\
  &= \frac{2\mu}{\Gamma\!\left(\frac d2+2\right)}
  \left(\frac{m\mu}{2\pi}\right)^{d/2}.
 \end{split}
\end{equation}
The next-to-leading-order correction is $O(\bar\eps)$, which corresponds
to the mean-field correction
\begin{equation}
 \begin{split}
  P_2 &= \bar g^2
  \left[\int\!\frac{d\p}{(2\pi)^d}\theta(\mu-\ep)\right]^2 \\
  &= \frac{\bar\eps\,\mu}{\Gamma\!\left(\frac d2+1\right)^2}
  \left(\frac{m\mu}{2\pi}\right)^{d/2}.
 \end{split}
\end{equation}

\begin{figure}[tp]
 \includegraphics[width=0.40\textwidth,clip]{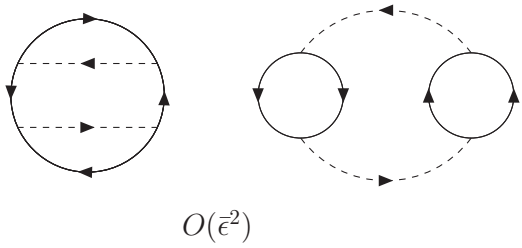}
 \caption{Vacuum diagrams contributing to the pressure to the
 next-to-next-to-leading order in $\bar\eps$.  In the right diagram, the
 counter vertex in Fig.~\ref{fig:cancel-2d}(b) for each bubble
 subdiagram is implicitly understood.  \label{fig:potential-2d}}
\end{figure}

To the next-to-next-to-leading order in $\bar\eps$, 
the pressure receives $O(\bar\eps^2)$ corrections from two three-loop
diagrams depicted in Fig.~\ref{fig:potential-2d}.  The left diagram is
easily evaluated as
\begin{align}
 P_{3a}
 &= \left[\bar g^2\int\!\frac{d\p}{(2\pi)^d}\theta(\mu-\ep)\right]^2
 \int\!\frac{d\q}{(2\pi)^d}\delta(\mu-\eq) \notag\\
 &= \frac{\bar\eps^2\mu}{\Gamma\!\left(\frac d2\right)
 \Gamma\!\left(\frac d2+1\right)^2}
 \left(\frac{m\mu}{2\pi}\right)^{d/2}. 
\end{align}
Now the right diagram in Fig.~\ref{fig:potential-2d} is written
as~\cite{Nishida:2006wk}
\begin{equation}\label{eq:P_3b}
 \begin{split}
  P_{3b} &= \bar g^2\int\!\frac{d\k d\p}{(2\pi)^{2d}}\,
  \theta(\mu-\varepsilon_{\p+\frac\k2})
  \theta(\mu-\varepsilon_{\p-\frac\k2}) \\
  &\quad\times \left[1+\bar g^2\!\int\!\frac{d\q}{(2\pi)^d}
  \frac{\theta(\varepsilon_{\q+\frac\k2}-\mu)
  \theta(\varepsilon_{\q-\frac\k2}-\mu)}{2\eq-2\ep}\right],
 \end{split}
\end{equation}
where the frequency integrations are already performed.  We note that
$+1$ in the square brackets comes from the counter vertex in
$\mathcal{L}_2$.  Due to the $\theta$ functions, the ranges of
integrations over $\ek$, $\ep$, and $\eq$ are limited to
$0\leq\ek\leq4\mu$, $0\leq\ep\leq\Lambda_p$, and $\Lambda_q\leq\eq$,
where
\begin{equation}
 \sqrt{\Lambda_p}=\frac{-|\cos\chi_p|\sqrt\ek+\sqrt{4\mu-\ek\sin^2\chi_p}}2
\end{equation}
and
\begin{equation}
  \sqrt{\Lambda_q}=\frac{|\cos\chi_q|\sqrt\ek+\sqrt{4\mu-\ek\sin^2\chi_q}}2,
\end{equation}
with $\cos\chi_p=\hat\k\cdot\hat\p$ and $\cos\chi_q=\hat\k\cdot\hat\q$.
The integration over $\eq$ can be performed analytically using
dimensional regularization.  As a result, the expression in the square
brackets in Eq.~(\ref{eq:P_3b}) becomes
\begin{equation}
 \bigl[\cdots\bigr] = 
 -\frac\gamma2\bar\eps - \frac{\bar\eps}2\int_0^\pi\!\frac{d\chi_q}{\pi}
  \ln\!\left(\frac{\Lambda_q-\ep}\mu\right)
  + O(\bar\eps^2).
\end{equation}
Then, introducing dimensionless variables $z=\ek/\mu$,
$\tilde\Lambda_{p(q)}=\Lambda_{p(q)}/\mu$ and 
performing the integration over $\ep/\mu$, we obtain the following
expression for $P_{3b}$:
\begin{equation}
 \begin{split}
  P_{3b} &= -\bar\eps^2\frac{m\mu^2}{2\pi}
  \biggl[\frac\gamma2 + \frac12\int_0^4\!dz\int_0^\pi\!\frac{d\chi_p}{\pi}
  \int_0^\pi\!\frac{d\chi_q}{\pi} \\
  &\qquad\times \left\{\tilde\Lambda_q\ln\tilde\Lambda_q
  -(\tilde\Lambda_q-\tilde\Lambda_p)\ln(\tilde\Lambda_q-\tilde\Lambda_p)
  -\tilde\Lambda_p\right\}\biggr].
 \end{split}
\end{equation}
Finally the numerical integrations over $z$, $\chi_p$, and $\chi_q$ lead
to
\begin{equation}
 P_{3b} = -\bar\eps^2\frac{m\mu^2}{2\pi}
  \left(\frac\gamma2 + 0.0568528\right) + O(\bar\eps^3).
\end{equation}

Consequently, we obtain the pressure up to the next-to-next-to-leading
order in $\bar\eps$ as
\begin{equation}
 \begin{split}
  P &= P_\mathrm{free} + P_2 + P_{3a} + P_{3b} \\
  &= P_\mathrm{free}
  \left[1 + \bar\eps + 0.6931472\,\bar\eps^2 + O(\bar\eps^3)\right].
 \end{split}
\end{equation}
The universal parameter of the unitary Fermi gas in Eq.~(\ref{eq:xi})
can be equivalently expressed as $\xi_d=\mu/\eF$.  From the
thermodynamic relationship $n=\d P/\d\mu$ and the definition of the
Fermi energy in $d$ spatial dimensions, 
\begin{equation}
 \eF=\frac{2\pi}m
  \left[\frac12\,\Gamma\!\left(\frac d2+1\right)n\right]^{2/d}, 
\end{equation}
we can determine $\xi_d$ from the $\bar\eps$ expansion to be
\begin{equation}
 \begin{split}
  \xi_{2+\bar\eps} &= \left[1 + \bar\eps 
  - 0.6931472\,\bar\eps^2\right]^{-2/(2+\bar\eps)} \\
  &= 1 - \bar\eps + 0.8068528\,\bar\eps^2 + O(\bar\eps^3).
 \end{split}
\end{equation}
This is the result shown in Eq.~(\ref{eq:xi_2d}).


\begin{thebibliography}{99}

\bibitem{review_theory}
  For recent reviews, see
%\bibitem{Bloch:2008}
  I.~Bloch, J.~Dalibard, and W.~Zwerger,
  %``Many-body physics with ultracold gases,''
  Rev.\ Mod.\ Phys.\ {\bf 80}, 885 (2008);
  %[arXiv:0704.3011 (cond-mat.other)].
%
%\bibitem{Giorgini:2008}
  S.~Giorgini, L.~P.~Pitaevskii, and S.~Stringari,
  %``Theory of ultracold atomic Fermi gases,''
  Rev.\ Mod.\ Phys.\ {\bf 80}, 1215 (2008).
  %[arXiv:0706.3360 (cond-mat.other)].

\bibitem{Ketterle:2008}
  W.~Ketterle and M.~W.~Zwierlein,
  %``Making, probing and understanding ultracold Fermi gases,''
  in Proceedings of the International School of Physics ``Enrico Fermi,''
  Varenna, 2006, edited by M.~Inguscio, W.~Ketterle, and C.~Salomon
  (IOS Press, Amsterdam, 2008),
  arXiv:0801.2500, %[cond-mat.other],
  and references therein.

\bibitem{Mehen:1999nd}
  T.~Mehen, I.~W.~Stewart, and M.~B.~Wise,
  %``Conformal invariance for non-relativistic field theory,''
  Phys.\ Lett.\ B {\bf 474}, 145 (2000).
  %[arXiv:hep-th/9910025].
  %%CITATION = PHLTA,B474,145;%%

\bibitem{Son:2005rv}
  D.~T.~Son and M.~Wingate,
  %``General coordinate invariance and conformal invariance in nonrelativistic
  %physics: Unitary Fermi gas,''
  Ann.\ Phys.\ (N.Y.) {\bf 321}, 197 (2006).
  %[arXiv:cond-mat/0509786].
  %%CITATION = APNYA,321,197;%%

\bibitem{Nishida:2007pj}
  Y.~Nishida and D.~T.~Son,
  %``Nonrelativistic conformal field theories,''
  Phys.\ Rev.\ D {\bf 76}, 086004 (2007).
  %[arXiv:0706.3746 (hep-th)].
  %%CITATION = PHRVA,D76,086004;%%

\bibitem{Mehen:2007dn}
  T.~Mehen,
  %``On Non-Relativistic Conformal Field Theory and Trapped Atoms: Virial
  %Theorems and the State-Operator Correspondence in Three Dimensions,''
  Phys.\ Rev.\ A {\bf 78}, 013614 (2008).
  %[arXiv:0712.0867 (cond-mat.other)].
  %%CITATION = ARXIV:0712.0867;%%

\bibitem{OHara:2002}
  K.~M.~O'Hara {\it et al.}, 
  %``Observation of a Strongly Interacting Degenerate Fermi Gas of Atoms,''
  Science {\bf 298}, 2179 (2002).

\bibitem{Bartenstein:2004}
  M.~Bartenstein {\it et al.},
  %``Crossover from a Molecular Bose-Einstein Condensate to a Degenerate Fermi Gas,''
  Phys.\ Rev.\ Lett.\ {\bf 92}, 120401 (2004).

\bibitem{Bourdel:2004}
  T.~Bourdel {\it et al.},
  %``Experimental Study of the BEC-BCS Crossover Region in Lithium 6,''
  Phys.\ Rev.\ Lett.\ {\bf 93}, 050401 (2004).

\bibitem{Kinast:2005}
  J.~Kinast {\it et al.},
  %``Heat Capacity of a Strongly Interacting Fermi Gas,''
  Science {\bf 307}, 1296 (2005).

\bibitem{Partridge:2006}
  G.~B.~Partridge {\it et al.},
  %``Pairing and Phase Separation in a Polarized Fermi Gas,''
  Science {\bf 311}, 503 (2006).

\bibitem{Stewart:2006}
  J.~T.~Stewart {\it et al.},
  %``Potential Energy of a {}^{40}K Fermi Gas in the BCS-BEC Crossover,''
  Phys.\ Rev.\ Lett.\ {\bf 97}, 220406 (2006).

\bibitem{Tarruell:2007}
  L.~Tarruell {\it et al.},
  %``Expansion of an ultra-cold lithium gas in the BEC-BCS crossover,''
  arXiv:cond-mat/0701181.

\bibitem{Carlson:2003}
  J.~Carlson, S.-Y.~Chang, V.~R.~Pandharipande, and K.~E.~Schmidt,
  %``Superfluid Fermi Gases with Large Scattering Length,''
  Phys.\ Rev.\ Lett.\ {\bf 91}, 050401 (2003).

\bibitem{Chang:2004}
  S.~Y.~Chang, V.~R.~Pandharipande, J.~Carlson, and K.~E.~Schmidt,
  %``Quantum Monte Carlo studies of superfluid Fermi gases,''
  Phys.\ Rev.\ A {\bf 70}, 043602 (2004).

\bibitem{Astrakharchik:2004}
  G.~E.~Astrakharchik, J.~Boronat, J.~Casulleras, and S.~Giorgini,
  %``Equation of State of a Fermi Gas in the BEC-BCS Crossover: A Quantum Monte Carlo Study''
  Phys.\ Rev.\ Lett.\ {\bf 93}, 200404 (2004).

\bibitem{Carlson:2005}
  J.~Carlson and S.~Reddy,
  %``Asymmetric Two-Component Fermion Systems in Strong Coupling,''
  Phys.\ Rev.\ Lett.\ {\bf 95}, 060401 (2005).

\bibitem{Bulgac:2006}
  A.~Bulgac, J.~E.~Drut, and P.~Magierski,
  %``Spin 1/2 Fermions in the Unitary Regime: A Superfluid of a New Type,''
  Phys.\ Rev.\ Lett.\ {\bf 96}, 090404 (2006).

\bibitem{Lee:2006}
  D.~Lee,
  %``Ground-state energy of spin-(1/2) fermions in the unitary limit,''
  Phys.\ Rev.\ B {\bf 73}, 115112 (2006).

\bibitem{Abe:2007}
  T.~Abe and R.~Seki,
  %``From Low-Density Neutron Matter to the Unitary Limit,''
  arXiv:0708.2524. %[nucl-th].

\bibitem{Bulgac:2008}
  A.~Bulgac, J.~E.~Drut, P.~Magierski, and G.~Wlazlowski,
  %``Gap and Pseudogap of a Unitary Fermi Gas by Quantum Monte Carlo,''
  arXiv:0801.1504. %[cond-mat.stat-mech].

\bibitem{Lee:2008}
  D.~Lee,
  %``The ground state energy at unitarity,''
  Phys.\ Rev.\ C {\bf 78}, 024001 (2008).
  %[arXiv:0803.1280 (nucl-th)].

\bibitem{Zhang}
  S.~Zhang, K.~E.~Schmidt, and J.~Carlson, unpublished.

\bibitem{Nishida:2006br}
  Y.~Nishida and D.~T.~Son,
  %``An epsilon expansion for Fermi gas at infinite scattering length,''
  Phys.\ Rev.\ Lett.\ {\bf 97}, 050403 (2006).
  %[arXiv:cond-mat/0604500].
  %%CITATION = PRLTA,97,050403;%%

\bibitem{Nishida:2006eu}
  Y.~Nishida and D.~T.~Son,
  %``Fermi gas near unitarity around four and two spatial dimensions,''
  Phys.\ Rev.\ A {\bf 75}, 063617 (2007).
  %[arXiv:cond-mat/0607835].
  %%CITATION = COND-MAT/0607835;%%

\bibitem{Nishida:2006rp}
  Y.~Nishida,
  %``Unitary Fermi gas at finite temperature in the epsilon expansion,''
  Phys.\ Rev.\ A {\bf 75}, 063618 (2007).
  %[arXiv:cond-mat/0608321].
  %%CITATION = COND-MAT/0608321;%%

\bibitem{Nussinov:2006}
  Z.~Nussinov and S.~Nussinov,
  %``Triviality of the BCS-BEC crossover in extended dimensions: Implications for the ground state energy,''
  Phys.\ Rev.\ A {\bf 74}, 053622 (2006).
  %[arXiv:cond-mat/0410597]

\bibitem{Arnold:2006fr}
  P.~Arnold, J.~E.~Drut, and D.~T.~Son,
  %``Next-to-next-to-leading-order epsilon expansion for a Fermi gas at
  %infinite scattering length,''
  Phys.\ Rev.\ A {\bf 75}, 043605 (2007).
  %[arXiv:cond-mat/0608477].
  %%CITATION = COND-MAT/0608477;%%

\bibitem{Nishida:2006wk}
  Y.~Nishida,
  %``Unitary Fermi gas in the epsilon expansion,''
  Ph.\,D.\ Thesis, University of Tokyo, 2007
  [available as arXiv:cond-mat/0703465].
  %%CITATION = COND-MAT/0703465;%%
  For the NNLO correction near $d=2$,
  see also the Appendix of the current paper.

\bibitem{Rupak:2006jj}
  G.~Rupak,
  %``Dimer scattering in the epsilon expansion,''
  arXiv:nucl-th/0605074.
  %%CITATION = NUCL-TH/0605074;%%

\bibitem{Rupak:2006et}
  G.~Rupak, T.~Schafer, and A.~Kryjevski,
  %``Polarized fermions in the unitarity limit,''
  Phys.\ Rev.\ A {\bf 75}, 023606 (2007).
  %[arXiv:cond-mat/0607834].
  %%CITATION = PHRVA,A75,023606;%%

\bibitem{Chen:2006wx}
  J.~W.~Chen and E.~Nakano,
  %``BEC-BCS Crossover in the Epsilon Expansion,''
  Phys.\ Rev.\ A {\bf 75}, 043620 (2007).
  %[arXiv:cond-mat/0610011].
  %%CITATION = PHRVA,A75,043620;%%

\bibitem{Kryjevski:2007au}
  A.~Kryjevski,
  %``Effective Lagrangian of unitary Fermi gas from $\varepsilon$ expansion,''
  Phys.\ Rev.\ A {\bf 78}, 043610 (2008);
  %arXiv:0712.2093 [nucl-th].
  %%CITATION = ARXIV:0712.2093;%%
%
%\bibitem{Kryjevski:2008si}
%  A.~Kryjevski,
  %``Low energy density correlation function of the degenerate unitary Fermi gas
  %from epsilon expansion,''
  arXiv:0804.2919. %[nucl-th].
  %%CITATION = ARXIV:0804.2919;%%

\bibitem{Rupak:2008xq}
  G.~Rupak and T.~Schafer,
  %``Density Functional Theory for non-relativistic Fermions in the Unitarity
  %Limit,''
  Nucl.\ Phys.\ {\bf A816}, 52 (2009).
  %arXiv:0804.2678 [nucl-th].
  %%CITATION = ARXIV:0804.2678;%%

\end{thebibliography}
\end{document}